\documentclass[12pt]{article}
\usepackage{amsfonts}
\usepackage{bm}
\usepackage{graphicx}
\usepackage{multicol}

\begin{document}

\title{\emph{}
Influence of noncommutativity on the motion of Sun-Earth-Moon system and the weak equivalence principle
}
\maketitle

\centerline {Kh. P. Gnatenko \footnote{E-Mail address: khrystyna.gnatenko@gmail.com},  V. M. Tkachuk \footnote{E-Mail address: voltkachuk@gmail.com}}
\medskip
\centerline {\small  $^{1,2}$ \it  Ivan Franko National University of Lviv, }
\centerline {\small \it Department for Theoretical Physics,12 Drahomanov St., Lviv, 79005, Ukraine}

\begin{abstract}
Features of motion of macroscopic body in gravitational field in a space with noncommutativity of coordinates and noncommutativity of momenta are considered in general case when coordinates and momenta of different particles satisfy noncommutative algebra with different parameters of noncommutativity. Influence of noncommutativity on the motion of three-body Sun-Earth-Moon system is examined. We show that because of noncommutativity the free fall accelerations of the Moon and the Earth toward the Sun in the case when the Moon and the Earth are at the same distance to the source of gravity  are not the same even if gravitational and inertial masses of the bodies are equal. Therefore, the Eotvos-parameter is not equal to zero and the weak equivalence principle is violated in noncommutative phase space. We estimate the corrections to the Eotvos-parameter caused by noncommutativity on the basis of Lunar laser ranging experiment results. We obtain that with high precision the ratio of parameter of momentum noncommutativity to mass is the same for different particles.


\end{abstract}

\section{Introduction}

In recent years much attention has been devoted to studies of quantized space realized on the basis of idea of noncommutativity. The idea was proposed by Heisenberg and later it was formulated by Snyder in his paper \cite{Snyder}.
The interest to the studies of noncommutative space is motivated by the development of String Theory and Quantum Gravity (see, for instance, \cite{Witten,Doplicher}).

In four dimensional  noncommutative phase space of canonical type  (2D configurational space and 2D momentum space) the commutation relations for coordinates and momenta are as follows
 \begin{eqnarray}
[X_{1},X_{2}]=i\hbar\theta,\label{form101}\\{}
[X_{i},P_{j}]=i\hbar\delta_{ij},\label{form1001}\\{}
[P_{1},P_{2}]=i\hbar\eta.\label{form10001}{}
\end{eqnarray}
Here $\theta$, $\eta$ are parameters noncommutativity which are constants, $i,j=(1,2)$.

Studies of physical systems in noncommutative space give a possibility to find influence of noncommutativity on their properties and to estimate the values of parameters of noncommutativity. Many physical problems were examined in the space, among them for example are harmonic oscillator  \cite{Hatzinikitas,Kijanka,Jing,Smailagic,Smailagic1,Giri,Geloun,Li14,Nath}  Landau problem \cite{Gamboa1,Horvathy,Dayi,Daszkiewicz1},  hydrogen atom \cite{Ho,Djemai,Bertolami,Chaichian,Chaichian1,Chair,Stern,Zaim2,Adorno,Khodja,Alavi,Masum,GnatenkoPLA14,GnatenkoIJMP17,GnatenkoConf}, gravitational quantum well \cite{Bertolami1,Bastos} classical systems with various potentials \cite{Gamboa,Romero,Mirza,Djemai1}, many-particle systems \cite{Ho,Djemai,Daszkiewicz,GnatenkoPLA13,GnatenkoJPS13,Daszkiewicz2,GnatenkoU16,GnatenkoIJ18} and many others.
At the same time it is important to study fundamental principles in noncommutative space, among them the weak equivalence principle.

The weak equivalence principle states that kinematic characteristics, such as velocity and position of a point mass in a gravitational field depend only on its initial position and velocity, and are independent of its mass, composition and structure. This principle is a restatement of the equality of gravitational and inertial masses.
  Implementation of this principle was considered in a space with noncommutativity of coordinates \cite{GnatenkoPLA13,GnatenkoJPS13,Saha,GnatenkoMPLA16} in a space with noncommutativity of coordinates and noncommutativity of momenta  \cite{Bastos1,Bertolami2}. It was shown that the equivalence principle is violated in a space with noncommutativity of coordinates and noncommutativity of momenta \cite{Bastos1}. In \cite{Bertolami2} the authors concluded that the equivalence principle holds in noncommutative phase space in the sense that an accelerated frame of reference is locally equivalent to
a gravitational field, unless noncommutative parameters are anisotropic ($\eta_{xy}\neq\eta_{xz}$). In our previous papers we proposed the ways to recover the weak equivalence principle in a space with noncommutativity of coordinates
 \cite{GnatenkoPLA13,GnatenkoMPLA16}, a space with noncommutativity of coordinates and noncommutativity of momenta \cite{GnatenkoPLA17}.

In the present paper we examine effect of noncommutativity on the motion of Sun-Earth-Moon system and consider the weak equivalence principle. We find influence of noncommutativity of coordinates and noncommutativity of momenta on the free fall accelerations of the Moon and the Earth toward the Sun. The results are compared with the results of the Lunar laser ranging experiment.

The paper is organized as follows. In Section 2 features of description of macroscopic body motion in noncommutative phase space are presented. The influence of noncommutativity on the motion of Sun-Earth-Moon system is studied in Section 3. In Section 4 the effect of noncommutativity of coordinates and noncommutativity of momenta on the free fall accelerations of the Moon and the Earth is obtained and the weak equivalence principle is examined. Conclusions are presented in Section 4.

\section{Features of description of macroscopic body motion in noncommutative phase space}

In general case different particles may feel noncommutativity with different parameters
\begin{eqnarray}
[X_{1}^{(a)},X_{2}^{(b)}]=i\hbar\delta^{ab}\theta_{a},\label{al0}\\{}
[X_{i}^{(a)},P_{j}^{(b)}]=i\hbar\delta^{ab}\delta_{ij},\\{}
[P_{1}^{(a)},P_{2}^{(b)}]=i\hbar\delta^{ab}\eta_{a},\label{al1}
\end{eqnarray}
here indexes $a$, $b$ label the particles, $\theta_{a}$, $\eta_{a}$ are parameters of coordinate and momentum noncommutativity. So, there is a problem of describing the motion of the center-of-mass of composite system in noncommutative phase space. This problem was studied in our paper \cite{GnatenkoPLA17}.

In the classical limit $\hbar\rightarrow0$ taking into account commutation relations (\ref{al0})-(\ref{al1}) one obtains the following Poisson brackets
\begin{eqnarray}
\{X_{1}^{(a)},X_{2}^{(b)}\}=\delta^{ab}\theta_{a},\label{p0}\\{}
\{X_{i}^{(a)},P_{j}^{(b)}\}=\delta^{ab}\delta_{ij},\\{}
\{P_{1}^{(a)},P_{2}^{(b)}\}=\delta^{ab}\eta_{a}.\label{p1}
\end{eqnarray}
 Defining momenta and coordinates of the center-of-mass of composite system, momenta and coordinates of relative motion in the traditional way
\begin{eqnarray}
\tilde{{\bf P}}=\sum_{a}{\bf P}^{(a)},\label{05}\\
\tilde{{\bf X}}=\sum_{a}\mu_{a}{\bf X}^{(a)},\label{00005}\\
\Delta{\bf P}^{{a}}={\bf P}^{(a)}-\mu_{a}\tilde{{\bf P}},\\
{\Delta\bf X}^{(a)}={\bf X}^{(a)}-\tilde{{\bf X}},\label{06}
\end{eqnarray}
with $\mu_a=m_{a}/M$, $M=\sum_{a}m_a$, one has
 \begin{eqnarray}
\{\tilde{X}_1,\tilde{X}_2\}=\tilde{\theta},\label{07}\\{}
\{\tilde{P}_1,\tilde{P}_2\}=\tilde{\eta},\\{}
\{\tilde{X}_i,\tilde{P}_j\}=\{\Delta{X}_i,\Delta{P}_j\}=\delta_{ij},\label{08}\\{}
\{\Delta{X}_1^{(a)},\Delta{X}_2^{(b)}\}=-\{\Delta{X}_2^{(a)},\Delta{X}_1^{(b)}\}=\delta^{ab}\theta_{a}-\mu_{a}\theta_{a}-\mu_{b}\theta_{b}+\tilde{\theta} ,\\{}
\{\Delta{P}_1^{(a)},\Delta{P}_2^{(b)}]=-\{\Delta{P}_2^{(a)},\Delta{P}_1^{(b)}\}=\delta^{ab}\eta_a-\mu_b\eta_a-\mu_a\eta_b+\mu_a\mu_b\tilde{\eta}.\\{}
 \{\tilde{X}_{1},\Delta X_{2}^{(a)}\}=\mu_{a}\theta_{a}-\tilde{\theta},\label{rel1}\\{}
 \{\tilde{P}_1,\Delta{P}^{a}_2\}=\eta_a-\mu_a\sum_{b}\eta_b.\label{rel2}
\end{eqnarray}
Here we take into account that coordinates $X_{i}^{(a)}$ and momenta $P_{i}^{(a)}$ satisfy (\ref{p0})-(\ref{p1}).
Parameters $\tilde{\theta}$, $\tilde{\eta}$ are effective parameters of coordinate noncommutativity and momentum noncommutativity  which describe the motion of the center-of-mass of composite system (macroscopic body). They are defined as
\begin{eqnarray}
\tilde{\theta}=\frac{\sum_{a}m_{a}^{2}\theta_{a}}{(\sum_{b}m_{b})^{2}},\label{eff}\\
\tilde{\eta}=\sum_{a}\eta_a.\label{eff2}
\end{eqnarray}

Note that the effective parameters of noncommutativity depend on the composition of a system \cite{GnatenkoPLA17}.

\section{Sun-Earth-Moon system in noncommutative phase space}

Let us study influence of noncommutativity on the motion of the Earth and the Moon in the gravitational field of the Sun. We consider the following Hamiltonian
\begin{eqnarray}
 H=\frac{({\bf P}^E)^{2}}{2m_E}+\frac{({\bf P}^M)^{2}}{2m_M}-G\frac{m_E m_S}{R_{ES}}-G\frac{m_M m_S}{R_{MS}}-G\frac{m_M m_E}{R_{EM}},
 \label{form12.5}
 \end{eqnarray}
here $m_S$, $m_E$, $m_M$ are the masses of Sun, Earth and Moon, respectively, $G$ is the gravitational constant.  Writing Hamiltonian (\ref{form12.5}) we suppose that influence of relative motion of particles which form the macroscopic body on the motion of its center-of-mass is not significant.

Choosing the Sun to be at the origin of the coordinate system we have
\begin{eqnarray}
R_{ES}=\sqrt{(X^E_1)^2+(X^E_2)^2},\\
R_{MS}=\sqrt{(X^M_1)^2+(X^M_2)^2},\\
R_{EM}=\sqrt{(X^E_1-X^M_1)^2+(X^E_2-X^M_2)^2},
 \end{eqnarray}
 where $X^E_i$, $X^M_i$ are coordinates of the Earth and the Moon, respectively, $i=(1,2)$.
 These coordinates and the momenta satisfy the following relations
 \begin{eqnarray}
 \{X_1^E,X_2^E\}=\theta_E, \ \ \{P^E_1,P^E_2\}=\eta_E,\ \ \{X^E_i,P^E_j\}=\delta_{ij},\label{dp}\\
 \{X_1^M,X_2^M\}=\theta_M, \ \ \{P^M_1,P^M_2\}=\eta_M,\ \ \{X^M_i,P^M_j\}=\delta_{ij},\\
 \{X_i^M,X_j^E\}= \{P_i^M,P_j^E\}=0  \label{dp1}
 \end{eqnarray}

Taking into account (\ref{dp})-(\ref{dp1}) the equations of motion read
\begin{eqnarray}
\dot{X}^E_1=\frac{P^E_{1}}{m_E}+\theta_E\frac{Gm_Em_S X^E_2}{R_{ES}^{3}}+\theta_E\frac{Gm_Em_M (X^E_2-X^M_2)}{R_{EM}^{3}},\label{form020}\\
\dot{X}^E_2=\frac{P^E_{2}}{m_E}-\theta_E\frac{Gm_Em_S X^E_1}{R_{ES}^{3}}-\theta_E\frac{Gm_Em_M (X^E_1-X^M_1)}{R_{EM}^{3}},\label{form021}\\
\dot{P}^E_1=\eta_E\frac{P^E_{2}}{m_E}-\frac{Gm_Em_S X^E_1}{R_{ES}^{3}}-\frac{Gm_Em_M (X^E_1-X^M_1)}{R_{EM}^{3}},\label{form022}\\
\dot{P}^E_2=-\eta_E\frac{P^E_{1}}{m_E}-\frac{Gm_Em_S X^E_2}{R_{ES}^{3}}-\frac{Gm_Em_M (X^E_2-X^M_2)}{R_{EM}^{3}},\label{form022}\\
\dot{X}^M_1=\frac{P^M_{1}}{m_M}+\theta_M\frac{Gm_M m_S X^M_2}{R_{MS}^{3}}-\theta_M\frac{G m_E m_M (X^E_2-X^M_2)}{R_{EM}^{3}},\label{form023}\\
\dot{X}^M_2=\frac{P^M_{2}}{m_M}-\theta_M\frac{Gm_M m_S X^E_1}{R_{MS}^{3}}+\theta_M\frac{G m_E m_M (X^E_1-X^M_1)}{R_{EM}^{3}},\label{form024}\\
\dot{P}^M_1=\eta_M\frac{P^M_{2}}{m_M}-\frac{Gm_M m_S X^M_1}{R_{MS}^{3}}+\frac{Gm_Em_M (X^E_1-X^M_1)}{R_{EM}^{3}},\label{form025}\\
\dot{P}^M_2=-\eta_M\frac{P^M_{1}}{m_M}-\frac{Gm_M m_S X^M_2}{R_{MS}^{3}}+\frac{Gm_Em_M (X^E_2-X^M_2)}{R_{EM}^{3}},\label{form026}
\end{eqnarray}

It is worth mentioning that because of the terms caused by noncommutativity in (\ref{form020})-(\ref{form026})  the velocity of macroscopic body in gravitational field depends on its mass. Also, taking into account definition of effective parameter of noncommutativity  (\ref{eff}) which correspond to motion of the center-of-mass of macroscopic body in noncommutative phase space,  we can state that the velocities of Earth and Moon depend on the composition of these bodies. From this follows that the weak equivalence principle is violated in noncommutative phase space.

\section{Estimation for effect of noncommutativity on the weak equivalence principle }
Stringent limit on any violation of the equivalence principle was provided by the Lunar laser ranging experiment \cite{Williams}.
The result was obtained on the basis of comparison of the free fall accelerations of the Earth and the Moon toward the Sun.
 According to the experiment the equivalence principle holds with accuracy
\begin{eqnarray}
\frac{\Delta a}{a}=\frac{2(a_E-a_M)}{a_E+a_M}=(-0.8\pm 1.3)\cdot10^{-13},\label{d}
\end{eqnarray}
where $a_E$, $a_M$ are the free fall accelerations of Earth and Moon toward the Sun when they are at the same distance from the Sun.
Let us use this result for analysis of the weak equivalence  principle in noncommutative phase space.

Using equations (\ref{form020})-(\ref{form026})  we can write  expressions for accelerations of the Earth and the Moon.
Up to the first order in the parameters of noncommutativity $\theta_M$, $\eta_M$, $\theta_E$, $\eta_E$ we have
\begin{eqnarray}
\ddot{X}^E_1=-\frac{Gm_S X^E_1}{{ R}_{ES}^{3}}-\frac{Gm_M (X^E_1-X^M_1)}{R_{EM}^{3}}+\eta_E\frac{\dot{X}^E_{2}}{m_E} +{\theta_E}\frac{Gm_{S}m_E \dot{X}^E_{2}}{R_{ES}^{3}}+\nonumber\\+{\theta_E}\frac{Gm_{M}m_E }{R_{EM}^{3}}(\dot{X}^E_{2}-\dot{X}^M_{2})
-\theta_E\frac{3Gm_S m_E}{R_{ES}^5}({\bf R}_{ES}\cdot{\bf \dot{R}}_{ES}){X^E_2}-\nonumber\\-\theta_E\frac{3Gm_M m_E}{R_{EM}^5}({\bf R}_{EM}\cdot{\bf \dot{R}}_{EM})(X^E_2-X^M_2),\nonumber\\ \label{ddot1E}\\
\ddot{X}^M_1=-\frac{Gm_S X^M_1}{{ R}_{MS}^{3}}+\frac{Gm_E (X^E_1-X^M_1)}{R_{EM}^{3}}+\eta_M\frac{\dot{X}^M_{2}}{m_M} +{\theta_M}\frac{Gm_{S}m_M \dot{X}^M_{2}}{R_{MS}^{3}}-\nonumber\\-{\theta_M}\frac{Gm_{M}m_E }{R_{EM}^{3}}(\dot{X}^E_{2}-\dot{X}^M_{2})
-\theta_M\frac{3Gm_S m_M}{R_{MS}^5}({\bf R}_{MS}\cdot{\bf \dot{R}}_{MS}){X^M_2}+\nonumber\\+\theta_M\frac{3Gm_M m_E}{R_{EM}^5}({\bf R}_{EM}\cdot{\bf \dot{R}}_{EM})(X^E_2-X^M_2).\nonumber\\ \label{ddot1M}
\end{eqnarray}
here ${\bf R}_{ES}(X_1^E,X_2^E)$, ${\bf R}_{MS}(X_1^M,X_2^M)$, ${\bf R}_{EM}(X_1^E-X_1^M,X_2^E-X_2^M)$.

Let us compare accelerations of the Moon and the Earth toward the Sun in the case when the Moon and the Earth are at the same distance to the source of gravity, $R_{MS}=R_{ES}=R$. It is convenient to choose the frame of references with $X_1$ axis perpendicular to the ${\bf R}_{EM}$ (passing through the middle of ${\bf R}_{EM}$), $X_2$ axis parallel to the ${\bf R}_{EM}$ and with origin at the Sun's center. Namely, $X^E_1=X^M_1=R\sqrt{1-R^2_{EM}/4R^2}\simeq R$ (here we take into account that $R_{EM}/R\sim10^{-3}$), and $X^E_2=-X^M_2=R_{EM}/2$. So, we can write the following expressions for the free fall accelerations of the Moon and the Earth toward the Sun
\begin{eqnarray}
a_E=\ddot{X}^E_1=-\frac{Gm_S}{R^{2}}+\eta_E\frac{\upsilon_E}{m_E} +{\theta_E}\frac{Gm_{S}m_E \upsilon_E}{R^{3}}\left(1-\frac{3R_{EM}}{2\upsilon_E R^2}({\bf R}_{ES}\cdot{\bf \dot{R}}_{ES})\right),\nonumber\\\label{ddot1E}\\
a_M=\ddot{X}^M_1=-\frac{Gm_S}{R^{2}}+\eta_M\frac{\upsilon_E}{m_M} +{\theta_M}\frac{Gm_{S}m_M \upsilon_E}{R^{3}}\left(1+\frac{3R_{EM}}{2\upsilon_E R^2}({\bf R}_{MS}\cdot{\bf \dot{R}}_{MS})\right),\nonumber\\\label{ddot1M}
\end{eqnarray}
here we take into account that $\dot{X}^E_{2}=\dot{X}^M_{2}=\upsilon_E$ ($\upsilon_E$ is the Earth orbital velocity). Also, we have  $\dot{X}^E_1=0$ and $\dot{X}^M_1=\upsilon_M$, where $\upsilon_E$ is the Moon orbital velocity. Note, that $R_{EM}/R\sim10^{-3}$, $\upsilon_M/\upsilon_E\sim10^{-2}$. So, the last terms in (\ref{ddot1E}), (\ref{ddot1M}) can be neglected, ${3R_{EM}({\bf R}_{ES}\cdot{\bf \dot{R}}_{ES})}/{2\upsilon_E R^2}\sim10^{-6}$, ${3R_{EM}({\bf R}_{MS}\cdot{\bf \dot{R}}_{MS})}/{2\upsilon_E R^2}\sim10^{-5}$.  The dimensionless Eotvos-parameter reads
\begin{eqnarray}
\frac{\Delta a}{a}=\frac{\Delta a^{\eta}}{a}+\frac{\Delta a^{\theta}}{a} \label{e}\\
\frac{\Delta a^{\eta}}{a}=\frac{\upsilon_ER^2}{Gm_S}\left(\frac{\eta_E}{m_E}-\frac{\eta_M}{m_M}\right),\label{de}\\
\frac{\Delta a^{\theta}}{a}=\frac{\upsilon_E}{R}\left(\theta_E m_E-\theta_M m_M\right).\label{dt}
\end{eqnarray}

Let us analyze the obtained result. Note that because of noncommutativity the Eotvos-parameter is not equal to zero even in the case of equality of gravitational and inertial masses of the bodies.  In (\ref{e}) one has term caused by the momentum noncommutativity ${\Delta a^{\eta}}/{a}$ which is proportional to $(\eta_E/m_E-\eta_M/m_M)$ and term caused by the noncommutativity of coordinates ${\Delta a^{\theta}}/{a}$ which is proportional to $(\theta_E m_E-\theta_M m_M)$. Parameters $\theta_E$, $\eta_E$,  $\theta_M$, $\eta_E$ are effective parameters of noncommutativity which are given by (\ref{eff}), (\ref{eff2}) and depend on the composition of the bodies. So, even if we consider as an example two bodies with the same masses but with different composition the Eotvos-parameter is not equal to zero.

In our paper \cite{GnatenkoPLA17} we proposed conditions on the parameters of noncommutativity on which the list of important results can be obtained in noncommutative phase space. Namely, we found that in the case when parameters of noncommutativity $\theta_i$, $\eta_i$ corresponding to a particle of mass $m_i$ satisfy relations
\begin{eqnarray}
m_i\theta_i=\gamma=const,\label{cond1}\\
\frac{\eta_i}{m_i}=\alpha=const,\label{cond2}
\end{eqnarray}
where $\gamma$, $\alpha$ are constants which do not depend on the mass, the weak equivalence principle is preserved; the kinetic energy has additivity property and does not depend on the composition; the Poisson brackets  (\ref{rel1}), (\ref{rel2}) are equal to zero therefore motion of the center-of-mass of composite system is independent on the relative motion; the noncommutative coordinates can be considered as kinematic variables \cite{GnatenkoMPLA17}; the effective parameters of noncommutativity $\tilde{\theta}$, $\tilde{\eta}$ describing the motion of the center-of-mass do not depend on its composition, one has
\begin{eqnarray}
\tilde{\theta}=\frac{\gamma}{M},\\
\tilde{\eta}=\alpha M,
\end{eqnarray}
where $M$ is the total mass of the system.

 Note, that in the case when conditions (\ref{cond1}), (\ref{cond2}) are satisfied the Eotvos-parameter (\ref{e}) is equal to zero and the equivalence principle is preserved.

If the conditions (\ref{cond1}), (\ref{cond2}) are not satisfied one has
\begin{eqnarray}
\frac{\eta_E}{m_E}=\alpha_E,\ \ \frac{\eta_M}{m_M}=\alpha_M,\\
\theta_E m_E=\gamma_E,\ \ \theta_M m_M=\gamma_M,
\end{eqnarray}
 here $\alpha_E\neq\alpha_M$, $\gamma_E\neq\gamma_M$. The result (\ref{e}) can be used to estimate the values of differences $\alpha_E-\alpha_M$, $\gamma_E-\gamma_M$. For this purpose we suppose that  effect of noncommutativity on motion of Earth and Moon which causes the violation of the weak equivalence principle is less than the experimental results for limits on violation of this principle.
So, we can write
\begin{eqnarray}
\left|\frac{\Delta a^{\theta}+\Delta a^{\eta}}{a}\right|\leq2.1\cdot10^{-13},
\end{eqnarray}
where $2.1\cdot10^{-13}$ is the largest value of $|\Delta a|/|a|$ obtained on the basis of the Lunar laser ranging experiment \cite{Williams}.
To estimate the orders of $\Delta \alpha=\alpha_E-\alpha_M$, $\Delta \gamma=\gamma_E-\gamma_M$ it is sufficiently to consider the following inequalities
\begin{eqnarray}
\left|\frac{\Delta a^{\theta}}{a}\right|\leq10^{-13},\label{nr}\\
\left|\frac{\Delta a^{\eta}}{a}\right|\leq10^{-13},\label{nr1}
\end{eqnarray}
Using (\ref{de}), (\ref{dt}) we obtain
\begin{eqnarray}
\Delta \alpha \leq10^{-20}\textrm{s}^{-1},\label{uupe}\\
\Delta \gamma \leq10^{-7}\textrm{s}.\label{uupte}
\end{eqnarray}
To analyze the obtained results the values of constants $\alpha$ and $\gamma$ have to be estimated. Stringent bound on the value of the parameter of momentum noncommutativity was found in \cite{Bertolami1} on the basis of studies of neutrons in gravitational quantum well in noncommutative phase space. The authors obtained  $\hbar|\eta|\leq2.4\times10^{-67}\textrm{kg}^2 \textrm{m}^2/\textrm{s}^2$. So, using this result we can write
\begin{eqnarray}
\alpha=\frac{\eta}{m_n}\leq10^{-6}\textrm{s}^{-1}.\label{alp}
\end{eqnarray}
here $m_n$ is the neutron mass.
Taking into account (\ref{uupe}), (\ref{alp}) we have
\begin{eqnarray}
\frac{\Delta \alpha}{\alpha}\leq10^{-14}.
\end{eqnarray}
So, we obtained quite strong restriction on the value of $\Delta \alpha$ and can conclude that proposed condition (\ref{cond2}) holds with high accuracy.

The constant $\gamma$ is of the order of $10^{-66}\textrm{s}$ \cite{GnatenkoPLA13}.  So, inequality (\ref{uupe}) does not impose strong restriction on the value of $\Delta \gamma$. The result is expectable because of reduction of the effective parameter of noncommutativity $\tilde{\theta}$ with respect to parameters corresponding to the individual particles (\ref{eff}). For instance, in particular case when a system is composed of $N$ identical particles of mass $m$ and parameters of noncommutativity $\theta$ from (\ref{eff}) we have $\tilde{\theta}=\theta/N$. So, the effect of coordinates noncommutativity on the properties of macroscopic systems is less than effect of the noncommutativity on the motion of individual particles. Therefore an experimental data with very hight accuracy are needed to obtain strong upper bound on the parameter $\theta$ or on the value of $\Delta \gamma$  on the basis of studies of macroscopic bodies in noncommutative space.

\section{Conclusions}

Noncommutative phase space of canonical type has been considered. The influence of noncommutativity of coordinates and noncommutativity of momenta on the motion of the Sun-Earth-Moon system has been studied. We have found that the free fall accelerations of the Moon and the Earth toward the Sun in the case when the Moon and the Earth are at the same distance to the source of gravity  are not the same even in the case of equality of gravitational and inertial masses of the bodies. Therefore the Eotvos-parameter  is not equal to zero (\ref{e}) and the equivalence principle is violated. The parameter depends on the values of $(\eta_E/m_E-\eta_M/m_M)$ and $(\theta_E m_E-\theta_M m_M)$.

We have used result for the Eotvos-parameter (\ref{e}) to estimate the values $(\eta_E/m_E-\eta_M/m_M)$ and $(\theta_E m_E-\theta_M m_M)$, namely to estimate the difference of constants $\alpha$ and $\gamma$ for Earth and Moon. For this purpose the data of the Lunar laser ranging experiment have been considered. Assuming that the effects of noncommutativity which cause the violation of the weak equivalence principle are less than the limits for any violation of the principle we have obtained upper bounds for the values $\Delta \alpha$ and $\Delta \gamma$ (\ref{uupe}), (\ref{uupte}).  The upper bound on $\Delta \alpha$ (\ref{uupe}) is quite stringent. The obtained restriction on $\Delta\gamma$ is not strong (\ref{uupte}). We have concluded that to find more strong restriction on the $\Delta \gamma$ more hight precision of experimental results is needed. This is because of reduction of effective parameter of coordinate noncommutativity with respect of increasing of number of particles in a system.

It is important to note that the Eotvos-parameter is equal to zero and the equivalence principle is recovered in noncommutative phase space when conditions (\ref{cond1}), (\ref{cond2}) are satisfied. The importance of these conditions is stressed by the number of results which can be obtained in noncommutative phase space in the case when they hold. Among them are preserving of the properties of the kinetic energy, independence of the motion of the center-of-mass on the relative motion \cite{GnatenkoPLA17,GnatenkoMPLA17}.  On the basis of our result for $\Delta \alpha$ (\ref{uupe}) we can conclude that the condition on the parameter of momentum noncommutativity (\ref{cond1}) holds with high precision, with high precision the ratio $\eta/m$ is the same for different particles.

\section{Acknowledgments}
This work was partly supported by the projects $\Phi\Phi$-63Hp (No. 0117U007190) and $\Phi\Phi$-30$\Phi$ (No. 0116U001539) from the Ministry of Education and Science of Ukraine.

\end{document}